\documentclass[floatfix, plb, nopacs, twocolumn,showkeys, preprintnumbers, nofootinbib, superscriptaddress]{revtex4-1}
\usepackage[utf8]{inputenc}
\usepackage[sort&compress]{natbib}
\usepackage{ulem}
\usepackage{overpic}
\usepackage{bm}
\usepackage{times}
\usepackage{amssymb,amsbsy,amsmath,amsfonts}
\usepackage{graphicx}
\usepackage{float}
\usepackage{color}
\usepackage{booktabs}
\usepackage{makecell}

\usepackage{rotating}
\usepackage{srcltx}
\usepackage{slashed}
\usepackage{subfigure}
\usepackage{multirow}
\usepackage{verbatim}
\usepackage{hyperref}
\usepackage{tabularx}

\DeclareUnicodeCharacter{3002}{HEREHEREHERE}

\begin{document}

\title{ One-proton emission from $^{148-151}$Lu in the DRHBc+WKB  approach}

\author{Yang Xiao}
\affiliation{School of Space and Environment, Beihang University, Beijing 102206, China}
\affiliation{School of Physics, Beihang University, Beijing 102206, China}

\author{Si-Zhe Xu}
\affiliation{School of Physics, Beihang University, Beijing 102206, China}

\author{Ru-You Zheng}
\affiliation{School of Physics, Beihang University, Beijing 102206, China}

\author{Xiang-Xiang Sun}
\affiliation{Institut für Kernphysik, Institute for Advanced Simulation and Jülich Center for Hadron Physics, Forschungszentrum Jülich, D-52425 Jülich, Germany}
\affiliation{School of Nuclear Science and Technology, University of Chinese Academy of Sciences, Beijing 100049, China}

\author{Li-Sheng Geng}\email{lisheng.geng@buaa.edu.cn}
\affiliation{School of Physics, Beihang University, Beijing 102206, China}
\affiliation{Peng Huanwu Collaborative Center for Research and Education, Beihang University, Beijing 100191, China}
\affiliation{Beijing Key Laboratory of Advanced Nuclear Materials and Physics, Beihang University, Beijing, 102206, China}
\affiliation{Southern Center for Nuclear-Science Theory (SCNT), Institute of Modern Physics, Chinese Academy of Sciences, Huizhou 516000, Guangdong Province, China}

\author{Shi-Sheng Zhang}\email{zss76@buaa.edu.cn}
\affiliation{School of Physics, Beihang University, Beijing 102206, China}

\date{\today}

\begin{abstract}
One-proton radioactivity in $^{149}$Lu, the latest identified proton emitter, is studied in the Wentzel-Kramers-Brillouin (WKB) approach with the proton-nucleus potential extracted from the deformed relativistic Hartree-Bogoliubov theory in continuum (DRHBc) for the first time. 
The  predicted half-life turns out to be consistent with the experimental measurement within uncertainties and (almost) independent of the density functionals in the DRHBc theory.
Such a microscopic self-consistent calculation reveals that  $^{149}$Lu is oblately deformed with a quadrupole deformation $\beta_2\approx -0.18$, and rules out the possibility of a prolate quadrupole deformation suggested in the nonadiabatic quasiparticle model. 
We also check the validity of this approach in the description of $^{150, 151}$Lu and their isomeric states. 
The deviations of the predicted half-lives from their experimental counterparts are mostly smaller than those of the theoretical studies without considering deformation effects. 
Furthermore, we predict $^{148}$Lu to be a more oblately deformed proton-emitter with a longer half-life than that of $^{149}$Lu, which can be checked in the future. Our studies show that
the DRHBc + WKB approach provides a new alternative  method to evaluate the half-lives of well-deformed proton emitters.
\end{abstract}


\maketitle

\section{Introduction}
\label{sec:Introduction}  
The exact location of the proton dripline and decay patterns of those nuclei close to the dripline are of great theoretical and experimental interests.
The half-lives of proton-emitters are sensitive to the decay energy, the orbital angular momentum of the emitted proton, and the spectroscopic information of the daughter nucleus~\cite{DELION2006113}. 
Therefore, it is challenging  to precisely predict their lifetimes and sometimes even the order of magnitudes. 

The conception of proton radioactivity was proposed even earlier than the discovery of proton~\cite{doi:10.1080/14786440808635390}. The first  proton emitter was observed and confirmed in an isomeric state of $^{53}$Co in 1970~\cite{Cerny:1970zvr,Jackson:1970wid,Cerny:1972gum}. 
Since then, more than 30 proton emitters have been discovered ~\cite{Woods:1997cs,BLANK2008403,Pfutzner:2011ju, PhysRevLett.128.112501}. 
Up to now, the properties of these proton emitters have been investigated in both macroscopic and microscopic methods, for reviews see, e.g., Refs.~\cite{BLANK2008403, Qi:2018idv}. 
Lately, a novel ground-state proton emitter $^{149}$Lu was discovered with the shortest directly measured half-life of $T_{1/2} = 450^{+170}_{-100}$ ns~\cite{PhysRevLett.128.112501}. It is the purpose of the present work to study the half-life of this nucleus and explore how its shape, {\it{i.e.}}, deformation, affects the result in a microscopic theory.

There are various ways to evaluate the half-lives of proton emitters~\cite{Woods:1997cs, Davids:1997zz, Aberg:1997kfq, Maglione:1998zz, PhysRevC.59.R2339, Balasubramaniam:2005cf, Basu:2005tf, PhysRevLett.96.072501, Bhattacharya:2007sd, Dong:2009hr,  Zhang_2010, Routray:2009qu, Qian_2010, Ferreira:2011zza, Zhao:2014lca, Qian:2016teg, Lim:2016gwi, Zdeb:2016cpp, PhysRevC.96.034619, Chen:2019jnc, Chen:2019aio, Cheng:2020hez,  Cheng:2020rob,Oishi:2022ued}, including perturbative approaches, such as the distorted wave Born approximation (DWBA)~\cite{ Aberg:1997kfq}, and the modified two potential approach (TPA)~\cite{ Aberg:1997kfq, Qian:2016teg, Cheng:2020hez, Cheng:2020rob}. 
Except those, the WKB approach is commonly used to calculate decay widths because of its simplicity and effectiveness~\cite{Woods:1997cs, Davids:1997zz, Aberg:1997kfq, PhysRevC.59.R2339, Balasubramaniam:2005cf, Basu:2005tf, PhysRevLett.96.072501, Bhattacharya:2007sd, Dong:2009hr,  Zhang_2010, Routray:2009qu, Qian_2010, Ferreira:2011zza, Zhao:2014lca, Lim:2016gwi, Zdeb:2016cpp, PhysRevC.96.034619, Chen:2019jnc, Chen:2019aio}.
Despite some conceptual differences between the WKB and perturbative approaches, the predicted half-lives based on the same nuclear potentials are rather similar, as shown  in Refs.~\cite{Aberg:1997kfq, 2022Xuthesis}. In particular, we note that the WKB results agrees well with the exact solutions for the widths of neutron resonant states of $^{17}$O, crucial for nucleosynthesis~\cite{2022EPJWC.26011037X}.  
Therefore, in the present work, we adopt the WKB approach  to study one-proton emission.

Up to now,  the short half-life of $^{149}$Lu has only been studied in  the nonadiabatic quasiparticle model~\cite{PhysRevLett.128.112501}. 
However,  its deformation cannot yet be completely determined. In Ref.~\cite{PhysRevLett.128.112501}, it was found that $^{149}$Lu could be either the most oblately deformed proton emitter ever observed with a quadrupole deformation $\beta_2 \lesssim -0.17$ or a prolately deformed proton emitter with a quadrupole deformation $\beta_2 \approx 0.15$. We note that most previous theoretical works predict an oblate shape for $^{149}$Lu~\cite{Lalazissis:1999bc, Moller:2015fba, Moller:2019jib, Geng:2003wb}. 

One of the most important inputs in theoretical studies of proton emissions is the nuclear potential. In a phenomenological model, although by adjusting free parameters of the model, a reasonable explanation of experimental half-lives can be achieved, the  uncertainties might be large because of the non-uniqueness of the nuclear potential in the case that  experimental measurements of nucleon levels are not available.
In that case, the predicting power of phenomenological model is rather limited.  
Therefore, microscopic potentials obtained in a self-consistent way are more reliable, which can account for various effects, such as deformations, pairing correlations, low-lying resonances and the continuum.
The deformed relativistic Hartree-Bogoliubov theory in continuum (DRHBc) theory~\cite{Zhou:2009sp, Li:2012gv} turns out to be one of the state of the art densify functional theories that can provide a microscopic nuclear potential. 

In our latest work on an unified description of the one-neutron halo, $^{31}$Ne, both its microscopic structure and the related reaction observables, we found that the DRHBc theory combined with the Glauber model can reproduce prominent halo evidences, such as the sudden increase of the reaction cross section of $^{31}$Ne bombarding $^{12}$C, and the narrow momentum distributions of the breakup interaction~\cite{Zhang:2021fpl,Zhong:2021yhm}. 
While these crucial features cannot be reproduced  using Skyrme forces within the non-relativistic mean field framework together with the Glauber model~\cite{Horiuchi:2012ca}. 
From the structure point of view, the DRHBc theory has successfully described halo nuclei~\cite{Zhou:2009sp, Li:2012gv, Meng:2015hta,  Sun:2018ekv, Zhang:2019qeu, Sun:2020tas, Yang:2021pbl, Sun:2021nyl, Sun:2021alk, Zhang:2023dhj} and shell evolution~\cite{Sun:2020tas} with deformation, pairing, and continuum effects properly taken into account
~\cite{In:2020asf, He:2021thz, DRHBcMassTable:2022uhi, DRHBcMassTable:2022rvn}.
These fruitful achievements trigger us to exploit the potential of the DRHBc theory to determine the deformation of the newly measured $^{149}$Lu, and provide structure inputs for the WKB approach to evaluate the half-life of the proton emitter $^{149}$Lu and those of its neighbouring nuclei.

This article is composed of three sections. Firstly, we briefly review the main formalism for the  DRHBc+WKB approach in Sect.~\ref{sec:Frame}.
In this approach, the half-lives, deformations, and spectroscopic factors of proton emitter $^{149}$Lu and its neighbouring nuclei are calculated and analyzed in Sect. ~\ref{sec:Results}. 
Finally, we provide a brief summary in Sect.~\ref{sec:summary}.

\section{Framework}
\label{sec:Frame} 
\subsection{The WKB approximation}
Inspired by our previous successful studies of the neutron decay widths~\cite{2022EPJWC.26011037X}, we adopt the simple and effective semi-classical WKB approximation to calculate the proton emission half-life, {\it{i.e.}},
\begin{align}\label{eq:halflife}
    T_{1/2} = \frac{\hbar \ln{2}}{S_F \it{\Gamma}}, 
\end{align}
where the spectroscopic factor $S_F$ for a nucleus with odd atomic number $Z$  in the
independent-quasiparticle approximation can be written as~\cite{Sorensen:1966zz}
\begin{align}\label{eq:Sfactor}
    S_F =u^2,
\end{align}
in which $u^2$ denotes the nonoccupation probability of the corresponding orbital in the daughter nucleus. In this work, the value of $S_F$ is obtained in the DRHBc framework self-consistently. 
The final decay width $\it{\Gamma}$ is obtained by averaging $\it{\Gamma}(\theta)$ in all directions with $\theta$ the orientation angle of the proton with respect to the symmetric axis of the daughter nucleus~\cite{Qian:2016teg}. 

In the WKB approximation, $\it{\Gamma}(\theta)$ is approximated  as
\begin{align}\label{eq:width}
    {\it{\Gamma}}(\theta) = N \frac{\hbar^2}{4\mu}\exp{\left(-2\int_{r_2}^{r_3}k(r,\theta) \text{d} r\right)},
\end{align}
in which $k(r,\theta)$ is related to the decay energy $E$ and potential barrier $V(r,\theta)$, {\it{i.e.}}, $k(r,\theta) = \sqrt{2\mu|E - V(r,\theta)|}$, $\mu$ refers to the reduced mass of the emitted proton and the daughter nucleus, $r$ denotes the distance between the proton and the mass center of the daughter nucleus, and $r_i$ corresponds to the $i$-th turning point at barrier $V(r,\theta)$ with respect to the decay energy $E$. The quasiclassic bound-state normalization factor $N$ in Eq.(\ref{eq:width}) is defined by~\cite{Buck:1992zza}
\begin{align}\label{eq:nor}
    N^{-1}(\theta) = \int_{r_1}^{r_2} \frac{\text{d}r}{k(r,\theta)} \cos^2 \left( \int_{r_1}^r \text{d} r' k(r',\theta)  -\frac{\pi}{4} \right).
\end{align}

The  potential barrier consists of three parts, {\it{i.e.}}, nuclear, Coulomb and centrifugal potentials. It reads
\begin{align}\label{eq:barrier}
    V(r,\theta) = V_N(r,\theta) + V_C(r,\theta) + \frac{\hbar^2}{2\mu} \frac{l(l+1)}{r^2},
\end{align}
where $V_{N/C}(r,\theta)$ refers to the nuclear/Coulomb potential respectively,  which is self-consistently calculated in the DRHBc theory.

\subsection{The DRHBc theory}
The potential $V(r,\theta)$ is crucial to calculate the half-life of a proton emitter. As stated above, we choose the DRHBc theory to provide the potential in a self-consistent way, in which the deformed RHB equations are solved with the Dirac Woods–Saxon basis~\cite{Zhou:2003jv} and the blocking effects are taken into account accordingly~\cite{Li:2012xaa}.
We briefly introduce the DRHBc theory in what follows. 
One can find more details  in Refs.~\cite{Li:2012gv,DRHBcMassTable:2020cfw} and references cited therein.

With the mean-field and no-sea approximations, the RHB equations for the nucleons treat the mean field and pairing correlations self-consistently, which read~\cite{peter1991}
\begin{align}\label{eq:RHB}
    \left(\begin{array}{cc}
        h_D - \lambda_\tau & \Delta \\
        -\Delta^* & -h_D^* + \lambda_\tau 
    \end{array}\right)
    \left(\begin{array}{c}
        U_k \\
        V_k
    \end{array}\right)
    = E_k \left(\begin{array}{c}
        U_k \\
        V_k
    \end{array}\right),
\end{align}
where $h_D$ is the Dirac Hamiltonian, $\lambda_\tau$ is the Fermi energy for protons/neutrons ($\tau = p/n$), $\Delta$ is the pairing potential, $E_k$ and $\left(U_k,V_k\right)^T$ refer to the quasiparticle energy and wave function, respectively. 
In coordinate space,  $h_D$ reads
\begin{align}\label{eq:hd}
    h_D (\bm{r})= \bm{\alpha}\cdot\bm{p} + V(\bm{r}) + \beta \left[M + S(\bm{r})\right],
\end{align}
where $V(\bm{r})/S(\bm{r})$ is the vector/scalar potential.
 The pairing potential $\Delta$ reads
\begin{align}\label{eq:pairing}
    \Delta(\bm{r_1},\bm{r_2}) = V^{pp}(\bm{r_1},\bm{r_2})\kappa (\bm{r_1},\bm{r_2}),
\end{align}
with $\kappa$ the pairing tensor. The pairing interaction in the particle-particle channel $V^{pp}$ takes the form of a density-dependent $\delta$ function
\begin{align}
    V^{pp} (\bm{r_1},\bm{r_2}) = V_0 \frac{1-P^{\sigma}}{2} \delta(\bm{r_1}-\bm{r_2})\left( 1-\frac{\rho(\bm{r_1})}{\rho_{\text{sat}}}\right),
\end{align}
where $V_0$ refers to the pairing strength, $\rho_{\text{sat}}$ denotes the saturation density of nuclear matter, and $\frac{1-P^{\sigma}}{2}$ is the projector into the spin $S= 0$ component. 

For axially deformed nuclei, the scalar potential, vector potential, and densities are expanded in terms of the Legendre polynomials~\cite{DRHBcMassTable:2022uhi}
\begin{align}
    f(\bm{r}) = \sum_\lambda f_\lambda(r)P_\lambda(\cos \theta),~~~\lambda = \text{even number}.
\end{align}
The single proton potential equals to the summation of vector potential $V(\bm{r})$ and scalar potentials $S(\bm{r})$, which can be obtained in a self-consistent manner. One can find the details in
Ref.~\cite{DRHBcMassTable:2020cfw} and references cited therein. 

\section{Results and discussion}
\label{sec:Results} 
\begin{figure}[hb]
\centering 
\includegraphics[height=8cm ,width=8cm,angle=0]{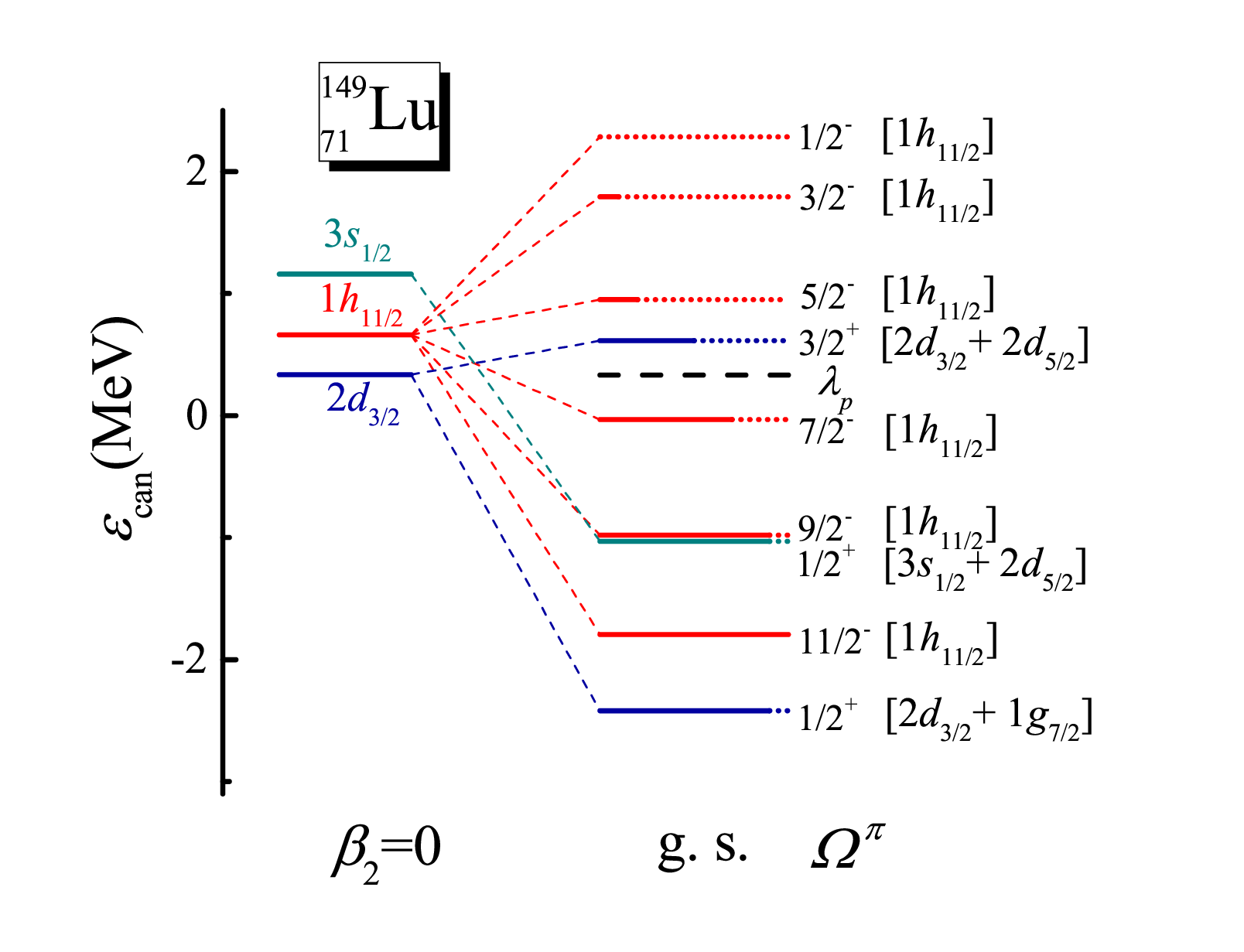}
\caption{Ground-state (g.s.) single-proton levels nearby the Fermi surface obtained in the DRHBc theory with the density functional PC-PK1. $\lambda_p$ denotes the Fermi energy. The length of the solid/dot lines represent the occupation/non-occupation probabilities of the single-particle orbitals. Quantum number  $\Omega^\pi$ and the main Woods-Saxon components are given for orbitals in the $sd$ shell and $h$ shell. }
\label{fig:spl}
\end{figure}

We calculate the half-life of proton emitter $^{149}$Lu in the WKB approximation with the proton-nucleus potential provided by the DRHBc theory. 
In this approach, three relativistic density functionals, namely PC-PK1~\cite{Zhao:2010hi}, NL3*~\cite{Lalazissis:2009zz}, and TMA~\cite{Sugahara:1993wz}, are selected for the study. 
Due to the similarity of the results, we take the density-functional PC-PK1 as an example to show in Fig.~\ref{fig:spl} the single-proton levels of $^{149}$Lu around the proton Fermi energy. 
A dominant $\pi h_{11/2}$ component for the ground-state wave-function of $^{149}$Lu is suggested in Ref.~\cite{PhysRevLett.128.112501}, which splits into six orbitals from ${1}/{2}^-$ to ${11}/{2}^-$ once the nucleus becomes deformed. 
The proton orbital is supposed to be blocked at orbitals $5/2^-$ and $7/2^-$ respectively, since both are consistent with the emission of a proton with $l_p = 5$ and their level energies are close to the proton Fermi energy.
The ground state of $^{149}$Lu is identified by minimizing the energy of the system. 

\begin{figure}[htpb]
\centering 
\includegraphics[height=8cm ,width=8cm,angle=0]{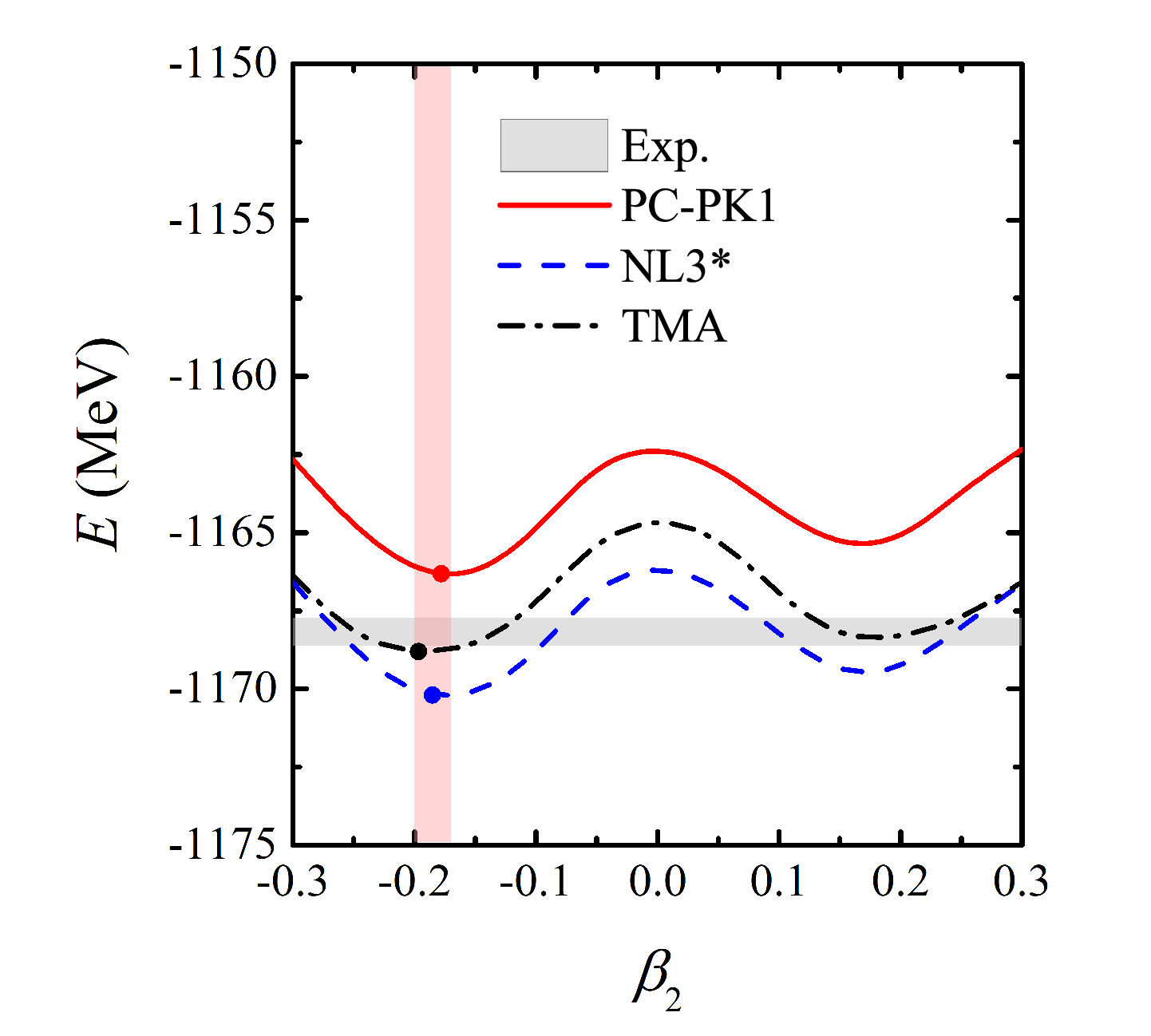}
\caption{Potential energy of $^{149}$Lu as a function of quadrupole deformation calculated in the DRHBc theory with PC-PK1, NL3*, and TMA density functional, respectively. The experimental energy range in gray is obtained from the binding energy of $^{148}$Yb from the NNDC database~\cite{citenndc} and the experimental decay energy of $^{149}$Lu~\cite{PhysRevLett.128.112501} taking into account experimental uncertainties. The red solid/ black dotdashed/ blue dashed lines represent the potential energies obtained with PC-PK1/ NL3*/ TMA density functionals. The band in the light red area covers the deformation of the ground states self-consistently obtained in the DRHBc theory.}
\label{fig:Ecm+Etotal}
\end{figure}

In Fig.~\ref{fig:Ecm+Etotal}, the potential energies of $^{149}$Lu as a function of the quadrupole deformation are displayed, which are obtained in the DRHBc theory with density functionals PC-PK1, NL3*, and TMA, respectively. 
One can see that the predicted quadrupole deformation of the ground state of $^{149}$Lu is relatively close, {\it{i.e.}}, $-0.178$ for PC-PK1, $-0.185$ for NL3*, and $-0.197$ for TMA, although the energies of the ground states obtained with the three density functionals are different.
Moreover, the minima of the potential energy curve support our
choice to block the valence proton at the orbital $5/2^-$ for all the three density functionals. We note that the second minimum is slightly higher than the first minimum by  about 0.5 MeV. 

To appreciate the fact that the DRHBc theory can successfully describe the ground-state properties of $^{149}$Lu, we list the nuclear structure inputs in Table~\ref{tb:inputs}. With the WKB approximation, the half-life of this proton emitter is computed and compared with the experimental measurement as well. 
It can be seen that $^{149}$Lu is found to be oblately deformed  in the DRHBc theory for all three density functionals, which agrees with one of the predictions of the nonadiabatic quasiparticle model, {\it{i.e.}}, $\beta_2\lesssim-0.17$. 
The calculated decay energy is  of the same order of the magnitude as  the experimental one.
Since the half-life of a proton emiter is very sensitive to the $Q$ value~\cite{BLANK2008403,DELION2006113,Qi:2018idv},
we utilize the experimental decay energy $Q$ in the calculations.
In this way, the half-life evaluated in the DRHBc + WKB approach is in good agreement with the experimental value at $1\sigma$ significance level, without any adjustable parameter. This indicates that the DRHBc theory correctly predicts the dominant contribution of the mean-field potential for $^{149}$Lu and the (non)occupation probability for $^{148}$Yb.

\begin{table}[h]
\centering
\caption{Half-life of  proton emitter $^{149}$Lu. Note that the experimental quadrupole deformation here refers to that of the nonadiabatic quasiparticle model instead of the experimental measurement as explained in Ref.~\cite{PhysRevLett.128.112501}.}
\label{tb:inputs}
\begin{tabular}{p{15mm}<{\centering}p{25mm}
<{\centering}p{13mm}<{\centering}p{13mm}<{\centering}p{13mm}<{\centering}}
\hline
\hline
\multirow{2}{*}{$^{149}$Lu} &\multirow{2}{*}{Exp.~\cite{PhysRevLett.128.112501}} &   \multicolumn{3}{c}{DRHBc (+ WKB)}   \\\cline{3-5}    
& & PC-PK1 & NL3* & TMA \\
\hline
Orbital & $1h_{{11/2}^-}$ & $5/2^-$& $5/2^-$& $5/2^-$\\
$Q$ (MeV) & 1.92(2) & 1.42 &1.16 & 1.22 \\ 
$S_F$ & / & 0.77 & 0.76 &0.68\\
$\beta_2$ & $\lesssim$ $-$ 0.17 or $\approx$ 0.15& $-$ 0.178 & $-$ 0.185 &$-$ 0.197\\
$T_{1/2}$ (ns) & $450_{-100}^{+170}$ & $538_{-131}^{+159}$& $469_{-110}^{+146}$& $544_{-128}^{+169}$\\
\hline\hline
\end{tabular}
\end{table}

\begin{figure}[htb]
\centering 
\includegraphics[height=8cm ,width=8cm,angle=0]{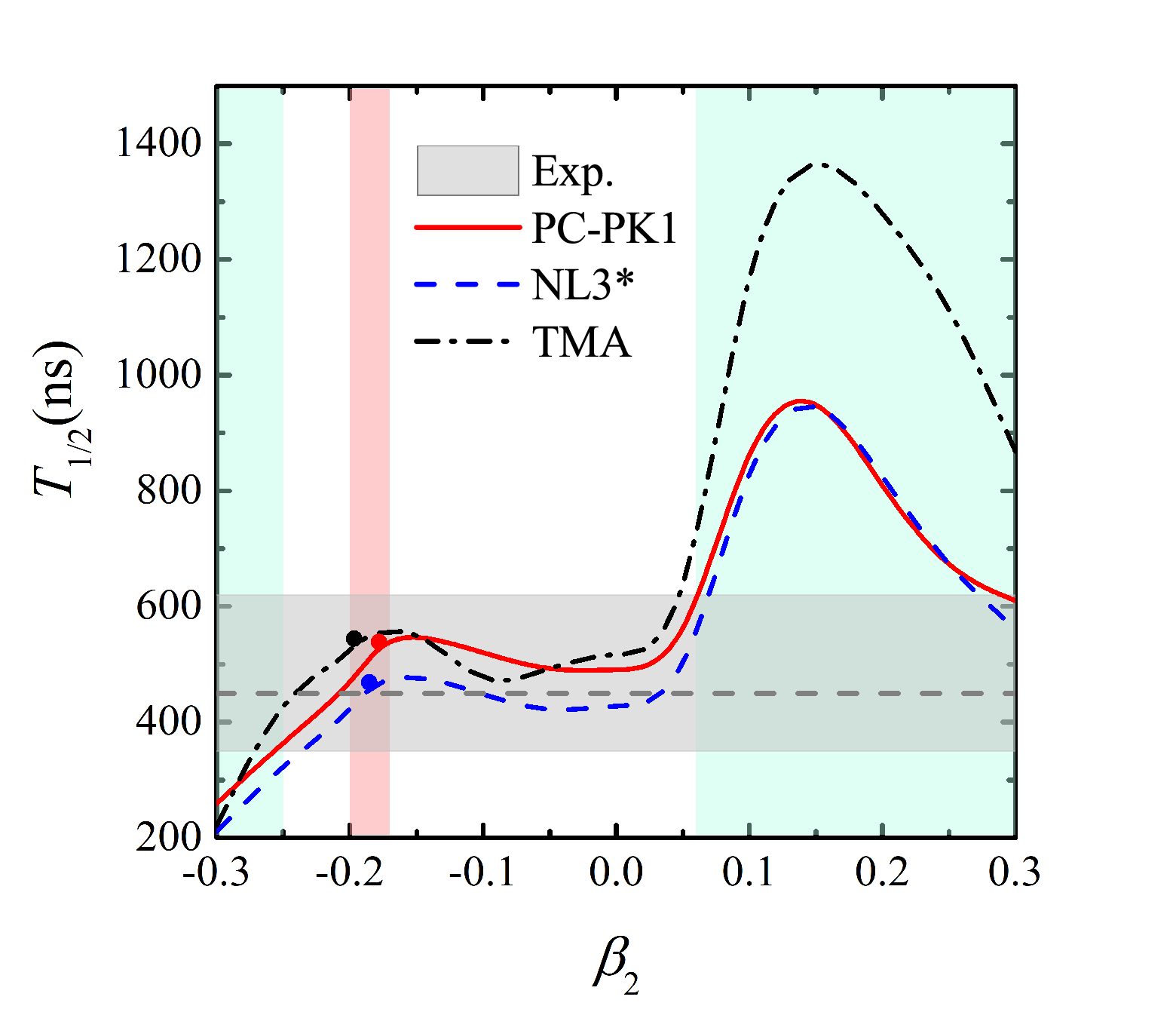}
\caption{Calculated and experimental half-life of $^{149}$Lu as a function of quadrupole deformation. The red solid/ black dot-dashed / blue dashed line represents the calculated half-life in the DRHBc theory with the  PC-PK1/ NL3*/ TMA density functional. The grey line denotes the central value of the experimental half-life and the shadow region represents the experimental error bar. The solid dots in the light red region correspond to the quadrupole deformations and half-lives of the ground-states obtained with three density functionals. The light green regions cover the values of the quadrupole deformations of $^{149}$Lu excluded in our calculations.}
\label{fig:halflife}
\end{figure}

In the nonadiabatic quasiparticle model, a prolate deformation of $\beta_2 \approx 0.15$~\cite{PhysRevLett.128.112501} was also recommended. 
In our model, the minimum at $\beta_2 \approx 0.15$ is the second minimum, about 0.5 MeV above the ground state as shown in Fig.~\ref{fig:Ecm+Etotal}.  
Therefore, we further study the half-life dependence on the deformation of $^{149}$Lu so that we can constrain the deformation by the measured half-life.
Our predictions in the DRHBc + WKB approach with the PC-PK1, NL3* and TMA density functionals are plotted in Fig.~\ref{fig:halflife}, respectively.
The calculated half-lives agree with the experimental data denoted by the grayed shadow region for the deformation range of $ -0.25 \lesssim \beta_2 \lesssim  0.06$, and are independent of the density functionals.
With the increase of the deformation $\beta_2$, our predictions exceed the gray region. 
We studied the rapid enhancement of half-life at $ \beta_2 \approx 0.1$ and found that it is caused by a switch of the orbit of emitted proton from ${5/2^-}$ to ${7/2^-}$. 
As a result, the spectroscopic factors shift from $0.77$, $0.76$, $0.68$ to $0.44$, $0.38$, $0.28$ for the PC-PK1, NL3* and TMA density functionals, respectively. Hence, the calculated half-lives suddenly increase according to Eq.~\eqref{eq:halflife}. 

We note that the calculated half-lives fall again within the experimental measurements as the quadrupole deformation $\beta_2$ increases to a relatively large value of $\sim 0.3$. However, such a large prolate deformation of $^{149}$Lu is not supported by the DRHBc theory because the ground state energy is much larger than the experimental value. 
Therefore, the prolate deformation of $\beta_2 > 0.06$, including $\beta_2 \approx 0.15$ predicted by the nonadiabatic quasiparticle model and $\beta_2 < - 0.25$ are out of the deformation range predicted
by the DRHBc theory, which are denoted in green and distinguished from the strongly recommended region in light red in Fig.~\ref{fig:halflife}. 
The deformation dependence of the half-life of $^{149}$Lu is still in debate and needs to be studied in other microscopic approaches and  checked by future experiments. 

As detailed above, the DRHBc + WKB approach is applicable to evaluate the half-life of $^{149}$Lu, which agrees well with the latest experimental measurement. We naturally expect that it also works for other lutetium proton-emitters, such as $^{150,151}$Lu and their isomers $^{150m,151m}$Lu.
If so, we will have the confidence to study other candidates, such as $^{148}$Lu.  
To study isomeric states, we take the same nuclear ground-state potentials~\cite{Lim:2016gwi}. 
In Table~\ref{tb:luisotope} we show the calculated half-lives of $^{150,151}$Lu and their isomers $^{150m,151m}$Lu, and predict the half-life of $^{148}$Lu in the  DRHBc + WKB approach. 
The results obtained in the spherical relativistic continuum Hartree-Bogoliubov (RCHB) theory + WKB approach~\cite{Lim:2016gwi} are also listed for comparison. 
It can be seen that the results from the DRHBc + WKB approach based on the PC-PK1 density functional are in general consistent with those from the RCHB + WKB calculations.
Define the deviation of theoretical predictions from the experimental data  $\chi$ by  $\chi = \left|\frac{T_{1/2}^{\text{theor.}}-T_{1/2}^{\text{exp.}}}{T_{1/2}^{\text{exp.}}}\right|$,
we can see from Table~\ref{tb:luisotope} that the $\chi$ obtained in the DRHBc + WKB approach is overall half of that obtained in the RCHB + WKB approach except for $^{150m}$Lu.
This highlights the importance of deformation effects to accurately describe the half-lives of the deformed proton emitters. 
The deviation between the DRHBc + WKB  result and the experimental value for $^{150m}$Lu may be due to the overestimated spectroscopic factor, which is reduced in ab initio calculations~\cite{Srivastava:2016bju}.     
For $^{148}$Lu, our study suggests that it is a more oblately deformed proton emitter with a half-life longer than that of $^{149}$Lu, which can be checked  by future experiments.
Although the NL3* density functional can better reproduce the half-life for the case of $^{149}$Lu which can be seen from Fig.~\ref{fig:halflife}, it is not as good as the PC-PK1 density functional in the overall description of the half-lives of  lutetium proton emitters, such as $^{150,150m,151,151m}$Lu.

\begin{table*}[htb!]
\small
\centering
\caption{Calculated and measured half-lives of lutetium proton emitters. The $\beta_2$ values are provided by the DRHBc theory. The experimental data are taken from Refs.~\cite{BLANK2008403,PhysRevLett.128.112501}. 
The RCHB results from Ref.~\cite{Lim:2016gwi} are shown for comparison. Both DRHBc and RCHB results are based on the PC-PK1 density functional. The $5/2^-$ orbital is a component of $1h_{11/2}$, while the $3/2^+$ orbital is a mixture of the configuration $2d_{3/2^+}$ and $2d_{5/2^+}$. The deviation $\chi$ is defined as $\chi = \left| \frac{T_{1/2}^{\text{theor.}} -T_{1/2}^{\text{exp.}}}  {T_{1/2}^{\text{exp.}}} \right| $.} 
\label{tb:luisotope}
\begin{tabular}{p{10mm}<{\centering}p{9mm}
<{\centering}p{9mm}<{\centering}p{9mm}<{\centering}p{9mm}<{\centering}p{12mm}<{\centering}p{9mm}<{\centering}p{9mm}<{\centering}p{10mm}<{\centering}p{10mm}<{\centering}p{15mm}<{\centering}p{20mm}<{\centering}p{10mm}<{\centering}p{9mm}<{\centering}p{9mm}<{\centering}}
\hline
\hline
\multirow{2}{*}{Nucleus} &\multirow{2}{*}{$\beta_2$} &   \multicolumn{3}{c}{Orbital} & \multicolumn{3}{c}{$Q$ (MeV)}  & \multicolumn{2}{c}{$S_F$} & \multicolumn{3}{c}{$T_{1/2}$ }& \multicolumn{2}{c}{{$\chi$} ($\%$) }\\\cline{3-15}  
 & &Exp.&DRHBc&RCHB&Exp.&DRHBc&RCHB&DRHBc&RCHB&Exp.&DRHBc&RCHB& DRHBc & RCHB\\
\hline
$^{148}$Lu &  $-0.184$ & / & $5/2^-$ & / &/ & 1.70&/ & 0.788&/&/&13.92$\mu$s & /&/&/\\
$^{149}$Lu & $-0.178$  & $1h_{11/2^-}$ &  $5/2^-$ &/ & 1.920(20)& 1.42&/&0.771 &/ &$450^{+170}_{-100}$ns & $538^{+159}_{-131}$ns &/& {19.56} &/\\
$^{150}$Lu & $-0.166$ & $1h_{11/2^-}$ & $5/2^-$  &$1h_{11/2^-}$ & 1.283(3)& 1.08 & 1.31 &0.753 &0.698&64.0(56)ms& $51.11^{+3.89}_{-7.58}$ms&88.46ms&{20.14}&{38.22}\\
$^{150m}$Lu & / &$2d_{3/2^+}$ & $3/2^+$ &$2d_{3/2^+}$ & 1.306(5)& /&/&0.692 &0.552&$43^{+7}_{-5}\mu$s &$27.34^{+3.40}_{-3.00}\mu$s&35.72$\mu$s &{36.42}&{16.93}\\
$^{151}$Lu & $-0.157$ & $1h_{11/2^-}$ &  $5/2^-$  &$1h_{11/2^-}$ & 1.253(3) & 0.82 &1.03&0.733&0.610&127.1(18)ms& $109.37^{+8.67}_{-7.95}$ms&176.38ms&{13.95}&{38.78}\\
$^{151m}$Lu & / & $2d_{3/2^+}$ & $3/2^+$ &$2d_{3/2^+}$ &1.332(10)& /&/&0.686&0.579&16(1)$\mu$s&$14.99^{+3.84}_{-3.03}\mu$s&18.03$\mu$s&{6.31}&{12.69}\\
\hline\hline
\end{tabular}
\end{table*}

\section{Summary and prospects}
\label{sec:summary}  
In summary, we have studied  the one-proton emission of the newly identified proton emitter $^{149}$Lu in the DRHBc + WKB approach. Using the self-consistently achieved proton-nucleus potential and spectroscopic factor obtained in the DRHBc theory with three different density functionals, we evaluated the half-life of $^{149}$Lu proton emission in the WKB approximation. 
Our theoretical predictions are in good agreement with the experiment data within uncertainties, {\it{i.e.}}, $538^{+159}_{-131}$ns, $469^{+146}_{-110}$ns, $544^{+169}_{-128}$ns for PC-PK1, NL3* and TMA density functionals respectively, which imply the robutness of our approach. Using the half-life as a constraint, we determine the quadrupole deformation of $^{149}$Lu, $\beta_2 \approx -0.18$. Our study rules out the possibility of a prolately deformed $^{149}$Lu suggested in the nonadiabatic quasiparticle model. This can be confirmed by future $\gamma$-ray spectroscopy experiments and checked by other theoretical works.

In the same approach, we also studied one-proton emission of other lutetium isotopes near $^{149}$Lu. It turns out that the results of the DRHBc+ WKB approach are on average better than those of the RCHB + WKB approach, which shows that the deformation effects play a relevant role in the study of proton-emission. Furthermore, we predicted $^{148}$Lu as a promising one-proton emitter with a oblate deformation $\beta_2 = - 0.184$, which can  be measured in the future. 

In summary, we showed that the DRHBc+WKB approach is an effective and efficient method to evaluate the half-lives of the newly discovered $^{149}$Lu and its neighbours. The sensitivity of the half-live to the deformation effects allows one to use the half-life to determine the deformation. Similar studies of  two-proton emission are in progress.

\section{Declaration of competing interest}
The authors declare that they have no known competing financial interests or personal relationships that could have appeared to
influence the work reported in this paper.

\section{Data availability}
Data will be made available on request.

\section{Acknowledgement} 
We would like to thank Dr. Yi-Bin Qian and  Dr. Kai-Yuan Zhang for valuable discussion. We also thank the DRHBc Mass Table collaboration for providing the DRHBc code and for useful suggestions. This work was partially supported by the National Natural Science Foundation of China
(Grant Nos. 12175010, 11975041, 11961141004, 1220530), the China Postdoctoral Science Foundation under Grant No. 2022M720360, the Deutsche Forschungsgemeinschaft
(DFG) and the National Natural Science Foundation of China (NSFC)
through the funds provided to the Sino-German Collaborative Research
Center TRR110 ``Symmetries and the Emergence of Structure in QCD''
(NSFC Grant No. 12070131001, DFG Project-ID 196253076). 

\bibliography{reference}

\end{document}